\documentstyle[12pt,epsf]{article}

\newcommand\jhep[3]{JHEP {\bf #1}, #3 (#2)} 
\newcommand\npb[3]{Nucl.\ Phys.\ B {\bf #1}, #3 (#2)} 
\newcommand\plb[3]{Phys.\ Lett.\ B {\bf #1}, #3 (#2)} 
\newcommand\Prd[3]{Phys.\ Rev.\ D {\bf #1}, #3 (#2)}
\newcommand\Prl[3]{Phys.\ Rev.\ Lett.\ {\bf #1}, #3 (#2)}
\newcommand\Rmp[3]{Rev.\ Mod.\ Phys.\ {\bf #1}, #3 (#2)}

\newcommand{\hepph}[1]{{\tt hep-ph/#1}}

\begin{document}

\begin{titlepage}

\begin{flushright}
SLAC-PUB-8231\\
hep-ph/9908404
\end{flushright}

\vspace{1.0cm}
\begin{center}
\Large\bf 
\boldmath
Large $\Delta I=\frac32$ Contribution to $\epsilon'/\epsilon$ in 
Supersymmetry
\unboldmath
\end{center}

\vspace{0.5cm}
\begin{center}
Alexander L. Kagan$^a$ and Matthias Neubert$^b$\\[0.1cm]
{\sl Stanford Linear Accelerator Center, Stanford Univerity\\
Stanford, California 94309, USA}
\end{center}

\vspace{0.5cm}
\begin{abstract}
\vspace{0.2cm}
\noindent  
We show that in supersymmetric extensions of the Standard Model 
gluino box diagrams can yield a large $\Delta I=\frac32$
contribution to $s\to d\bar q q$ FCNC processes, which may induce 
a sizable CP-violating contribution to the $I=2$ isospin amplitude 
in $K\to\pi\pi$ decays. This contribution only requires moderate
mass splitting between the right-handed squarks $\tilde u_R$ and 
$\tilde d_R$, and persists for squark masses of order 1\,TeV. 
Taking into account current bounds on $\mbox{Im}\,\delta_{sd}^{LL}$ 
from $K$--$\bar K$ mixing, the resulting contribution to 
$\epsilon'/\epsilon$ could be an order of magnitude larger than 
the measured value.
\end{abstract}

\vspace{0.5cm}
\centerline{(Submitted to Physical Review Letters)}

\vfil
\noindent
August 1999\\[0.7cm]
{\small
$^a$ On leave from: Department of Physics, University of Cincinnati, 
Cincinnati, Ohio 45221\\[0.1cm]
$^b$ On leave from: Newman Laboratory of Nuclear Studies, Cornell 
University, Ithaca, NY 14853}

\end{titlepage}

The recent confirmation of direct CP violation in $K\to\pi\pi$ 
decays is an important step in testing the 
Cabibbo--Kobayashi--Maskawa (CKM) mechanism for CP violation in 
the Standard Model. Combining the recent measurements by the KTeV 
and NA48 experiments \cite{new} with earlier results from NA31 and 
E731 \cite{old} gives $\mbox{Re}\,(\epsilon'/\epsilon)
=(2.12\pm 0.46)\times 10^{-3}$. This value tends to be higher than 
theoretical predictions in the Standard Model, which center below 
or around $1\times 10^{-3}$ \cite{updates}. Unfortunately, it is 
difficult to gauge the accuracy of these predictions, because they 
depend on hadronic matrix elements which at present cannot be 
computed from first principles. A Standard-Model explanation of 
$\epsilon'/\epsilon$ can therefore not be excluded. Nevertheless, 
it is interesting to ask how large $\epsilon'/\epsilon$ could be 
in extensions of the Standard Model.

In the context of supersymmetric models, it has been known for some 
time that it is possible to obtain a large contribution to 
$\epsilon'/\epsilon$ via the $\Delta I=\frac12$ chromomagnetic 
dipole operator without violating constraints from $K$--$\bar K$ 
mixing \cite{Gabb}. It has recently been pointed out that this 
mechanism can naturally be realized in various supersymmetric 
scenarios \cite{SUSY}. In this 
Letter we propose a new mechanism involving a supersymmetric 
contribution to $\epsilon'/\epsilon$ induced by $\Delta I=\frac32$ 
penguin operators. These operators are potentially important because 
their effect is enhanced by the $\Delta I=\frac12$ selection rule.
Unlike previous proposals, which involve left-right down-squark 
mass insertions, our effect relies on the left-left insertion 
$\delta_{sd}^{LL}$ and requires (moderate) isospin violation in
the right-handed squark sector.

The ratio $\epsilon'/\epsilon$ parametrizing the strength of direct 
CP violation in $K\to\pi\pi$ decays can be expressed as
\begin{equation}\label{eps}
   \frac{\epsilon'}{\epsilon}
   = i e^{i(\delta_2-\delta_0-\phi_\epsilon)}\,
   \frac{\omega}{\sqrt2\,|\epsilon|} \left( 
   \frac{\mbox{Im}\,A_2}{\mbox{Re}\,A_2}
   - \frac{\mbox{Im}\,A_0}{\mbox{Re}\,A_0} \right) ,
\end{equation}
where $A_I$ are the isospin amplitudes for the decays $K^0\to
(\pi\pi)_I$, $\delta_I$ are the corresponding strong-interaction 
phases, and the ratio $\omega=\mbox{Re}\,A_2/\mbox{Re}\,A_0\approx 
0.045$ signals the strong enhancement of $\Delta I=\frac12$ 
transitions over those with $\Delta I=\frac32$. From experiment, we 
take $|\epsilon|=(2.280\pm 0.013)\times 10^{-3}$ and 
$\phi_\epsilon=(43.49\pm 0.08)^\circ$ for the magnitude and phase 
of the parameter $\epsilon$ measuring CP violation in $K$--$\bar K$ 
mixing, and $\delta_2-\delta_0=-(42\pm 4)^\circ$ for the difference
of the S-wave $\pi\pi$ scattering phases in the two isospin
channels. It follows that, to an excellent approximation, 
$\epsilon'/\epsilon$ is real.

In the Standard Model, the isospin amplitudes $A_I$ receive small, 
CP-violating contributions via the ratio $(V_{ts}^* V_{td})/
(V_{us}^* V_{ud})$ of CKM matrix elements. This ratio enters 
through the interference of the $s\to u\bar u d$ tree diagram 
with penguin diagrams involving the exchange of a virtual top quark.
According to (\ref{eps}), contributions to $\epsilon'/\epsilon$ due 
to the $\Delta I=\frac32$ amplitude $\mbox{Im}\,A_2$ are enhanced 
relative to those due to the $\Delta I=\frac12$ amplitude 
$\mbox{Im}\,A_0$ by a factor of $\omega^{-1}\approx 22$. However, 
in the Standard Model the dominant CP-violating contributions to 
$\epsilon'/\epsilon$ are due to QCD penguin operators, which 
only contribute to $A_0$. Penguin contributions to $A_2$ arise
through electroweak interactions and are suppressed by a power 
of $\alpha$. Their effects on 
$\epsilon'/\epsilon$ are subleading and of the same order as 
isospin-violating effects such as $\pi^0$--$\eta$--$\eta'$ mixing.

Here we point out that in supersymmetric extensions of the Standard 
Model potentially large, CP-violating contributions can arise from 
flavor-changing {\em strong-interaction\/} processes induced by 
gluino box diagrams. Whereas in the limit of exact isospin symmetry
in the squark sector these graphs only induce $\Delta I=\frac12$ 
operators at low energies, in the presence of even moderate 
up-down squark mass splitting they can lead to operators with 
large $\Delta I=\frac32$ components. In the terminology of the 
standard effective weak Hamiltonian, this implies that the 
supersymmetric contributions to the Wilson coefficients of QCD and 
electroweak penguin operators can be of the same order. Specifically,
both sets of coefficients scale like $\alpha_s^2/\widetilde m^2$ with
$\widetilde m$ a generic supersymmetric mass, compared with 
$\alpha_s\alpha_W/m_W^2$ and $\alpha\alpha_W/m_W^2$, respectively, in 
the Standard Model. These contributions can be much larger than the 
electroweak penguins of the Standard Model even for supersymmetric 
masses of order 1\,TeV. On the other hand, supersymmetric contributions 
to the Wilson coefficients proportional to electroweak gauge couplings 
are parametrically suppressed and will not be considered here.

We find that the relevant $\Delta S=1$ gluino box diagrams lead to 
the effective Hamiltonian 
\[
   {\cal H}_{\rm eff} = \frac{G_F}{\sqrt2}
   \sum_{i=1}^4 \left[ c_i^q(\mu)\,Q_i^q(\mu)
   + \widetilde c_i^q(\mu)\,\widetilde Q_i^q(\mu) \right] 
   + \mbox{h.c.} \,,
\]
where 
\begin{eqnarray}
   Q_1^q &=& (\bar d_\alpha s_\alpha)_{V-A}\,
    (\bar q_\beta q_\beta)_{V+A} \,, \nonumber\\
   Q_2^q &=& (\bar d_\alpha s_\beta)_{V-A}\,
    (\bar q_\beta q_\alpha)_{V+A} \,, \nonumber\\
   Q_3^q &=& (\bar d_\alpha s_\alpha)_{V-A}\,
    (\bar q_\beta q_\beta)_{V-A} \,, \nonumber\\
   Q_4^q &=& (\bar d_\alpha s_\beta)_{V-A}\,
    (\bar q_\beta q_\alpha)_{V-A} \nonumber
\end{eqnarray}
are local four-quark operators renormalized at a scale 
$\mu\ll\widetilde m$, $\widetilde Q_i^q$ are operators of opposite
chirally obtained by interchanging $V-A\leftrightarrow V+A$, and a 
summation over $q=u,d,\dots$ and over color indices $\alpha,\beta$ 
is implied. In the calculation of the coefficient functions we use
the mass insertion approximation, in which case the 
gluino--quark--squark couplings are flavor diagonal. Flavor mixing
is due to small deviations from squark-mass degeneracy and is 
parametrized by dimensionless 
quantities $\delta_{ij}^{AB}$, where $i,j$ are squark flavor indices 
and $A,B$ refer to the chiralities of the corresponding quarks 
(see, e.g., \cite{Gabb}). In general, these mass 
insertions can carry new CP-violating phases. Contributions involving 
left-right squark mixing are neglected, since they are quadratic in 
small mass insertion parameters, i.e., proportional to 
$\delta_{sd}^{LR}\,\delta_{qq}^{LR}$. We define the dimensionless 
ratios
\[
   x_u^{L,R} = \left( \frac{m_{\tilde u_{L,R}}}{m_{\tilde g}}
   \right)^2 , \qquad 
   x_d^{L,R} = \left( \frac{m_{\tilde d_{L,R}}}{m_{\tilde g}}
   \right)^2 ,
\]
where $m_{\tilde u_{L,R}}$ and $m_{\tilde d_{L,R}}$ denote the 
average 
left- or right-handed squark masses in the up and down sector, 
respectively. SU(2)$_L$ gauge symmetry implies that the mass 
splitting between the left-handed up- and down-squarks
must be tiny; however, we will not assume such a degeneracy
between the right-handed squarks. For the Wilson coefficients 
$c_i^q$ at the supersymmetric matching scale $\widetilde m$
we then obtain
\begin{eqnarray}
   c_1^q &=& \frac{\alpha_s^2\delta_{sd}^{LL}}
                  {2\sqrt2 G_F m_{\tilde g}^2} 
    \left[ \frac{1}{18}\,f(x_d^L,x_q^R)
    - \frac{5}{18}\,g(x_d^L,x_q^R) \right] , \nonumber\\
   c_2^q &=& \frac{\alpha_s^2\delta_{sd}^{LL}}
                  {2\sqrt2 G_F m_{\tilde g}^2} 
    \left[ \frac{7}{6}\,f(x_d^L,x_q^R)
    + \frac{1}{6}\,g(x_d^L,x_q^R) \right] , \nonumber\\
   c_3^q &=& \frac{\alpha_s^2\delta_{sd}^{LL}}
                  {2\sqrt2 G_F m_{\tilde g}^2} 
    \left[ - \frac{5}{9}\,f(x_d^L,x_q^L)
    + \frac{1}{36}\,g(x_d^L,x_q^L) \right] , \nonumber\\
   c_4^q &=& \frac{\alpha_s^2\delta_{sd}^{LL}}
                  {2\sqrt2 G_F m_{\tilde g}^2} 
    \left[ \frac{1}{3}\,f(x_d^L,x_q^L)
    + \frac{7}{12}\,g(x_d^L,x_q^L) \right] , \nonumber
\end{eqnarray}
where 
\begin{eqnarray}
   f(x,y) &=& \frac{x(x+1-2y)}{(x-1)^2(y-1)(x-y)}
    - \frac{xy\ln y}{(y-1)^2(x-y)^2} 
    + \frac{x[2x^2-(x+1)y]\ln x}{(x-1)^3(x-y)^2} \,,
    \nonumber\\
   g(x,y) &=& \frac{x[-2x+(x+1)y]}{(x-1)^2(y-1)(x-y)}
    + \frac{xy^2\ln y}{(y-1)^2(x-y)^2} 
    - \frac{x^2[x(x+1)-2y]\ln x}{(x-1)^3(x-y)^2} \,. 
    \nonumber
\end{eqnarray}
The results for the coefficients $\widetilde c_i^q$ are obtained 
by interchanging $L\leftrightarrow R$ in the expressions for 
$c_i^q$. 

It is straightforward to relate the quantities $c_i^q$ to the 
Wilson coefficients appearing in the effective weak
Hamiltonian of the Standard Model 
as defined, e.g., in \cite{Heff}. We find 
($\lambda_t=V_{ts}^* V_{td}$)
\begin{eqnarray}
   (-\lambda_t)\,C_3 &=& \frac{c_3^u + 2c_3^d}{3} \,, \qquad
    (-\lambda_t)\,C_4 = \frac{c_4^u + 2c_4^d}{3} \,, \nonumber\\
   (-\lambda_t)\,C_5 &=& \frac{c_1^u + 2c_1^d}{3} \,, \qquad
    (-\lambda_t)\,C_6 = \frac{c_2^u + 2c_2^d}{3} \nonumber
\end{eqnarray}
for the QCD penguin coefficients, and
\begin{equation}\label{EWcs}
   \frac32(-\lambda_t)\,C_{i+6} = c_i^u - c_i^d 
   \equiv \Delta c_i \,; \quad i=1\dots 4
\end{equation}
for the coefficients of the electroweak penguin operators. 
The supersymmetric contributions to the 
electroweak penguin coefficients vanish in the limit of universal 
squark masses. However, for moderate up--down squark 
mass splitting the differences $\Delta c_i=c_i^u-c_i^d$ are of the 
same order as the coefficients
$c_i^q$ themselves. In this case gluino box contributions to
QCD and electroweak penguin operators are of similar magnitude. 
This conclusion is unaltered when additional contributions
to $C_{3\dots 6}$ from QCD penguin diagrams with gluino loops
are taken into account. Because the 
electroweak penguin operators contain $\Delta I=\frac32$ 
components their contributions to $\epsilon'/\epsilon$ are 
strongly enhanced and thus are expected to be an order of
magnitude larger than the contributions from
the QCD penguin operators. In this Letter, we focus only on 
these enhanced contributions. 

The renormalization-group evolution of the coefficients 
$\Delta c_i$ (and $\Delta\widetilde c_i$) from the supersymmetric 
matching scale $\widetilde m$ down to low energies is well known. 
In leading logarithmic approximation, one obtains \cite{Heff}
\begin{eqnarray}
   \Delta c_1(\mu) &=& \kappa^{-1/\beta_0}\,c_1(\widetilde m) \,,
    \nonumber\\
   \Delta c_2(\mu) + \frac{\Delta c_1(\mu)}{3}
   &=& \kappa^{8/\beta_0} \left[ \Delta c_2(\widetilde m)
    + \frac{\Delta c_1(\widetilde m)}{3} \right] , \nonumber\\
   \Delta c_3(\mu) + \Delta c_4(\mu)
   &=& \kappa^{-2/\beta_0} \left[ \Delta c_3(\widetilde m)
    + \Delta c_4(\widetilde m) \right] \,, \nonumber\\
   \Delta c_3(\mu) - \Delta c_4(\mu)
   &=& \kappa^{4/\beta_0} \left[ \Delta c_3(\widetilde m)
    - \Delta c_4(\widetilde m) \right] \,, \nonumber
\end{eqnarray}
where $\kappa=\alpha_s(\mu)/\alpha_s(\widetilde m)$, and 
$\beta_0=11-\frac23 n_f$. 
It is understood that the value of $\beta_0$ is changed
at each quark threshold.
We use the two-loop running coupling normalized to 
$\alpha_s(m_Z)=0.119$ and take the quark thresholds at 
$m_t=165\,$GeV, $m_b=4.25\,$GeV and $m_c=1.3\,$GeV. In 
Table~\ref{tab:1} we give the imaginary parts of the coefficients 
$\Delta c_{1,2}$ and $\Delta\widetilde c_{3,4}$ at the scale 
$\mu=m_c$, obtained for the illustrative choice 
$\widetilde m=m_{\tilde g}=m_{\tilde d_L}=m_{\tilde d_R}=500$\,GeV 
and three (larger) values of $m_{\tilde u_R}$. 
Since the mass 
splitting between the left-handed $\tilde u_L$ and $\tilde d_L$ 
squarks is tiny, we can safely 
neglect the coefficients $\Delta c_{3,4}$ and 
$\Delta\widetilde c_{1,2}$ in our analysis. Note that for fixed 
ratios of the supersymmetric masses the values of the coefficients
scale like $\widetilde m^{-2}$, i.e., significantly larger values
could be obtained for smaller masses. For comparison, the 
last column contains the imaginary parts of 
$\Delta c_{1\dots 4}$ in the Standard Model computed 
from (\ref{EWcs}) using $\mbox{Im}\,\lambda_t=1.2\times 10^{-4}$ 
and the next-to-leading order Wilson coefficients $C_i$ compiled
in \cite{Heff}. We observe that for 
supersymmetric masses of order 500\,GeV, and for a 
mass insertion parameter $\mbox{Im}\,\delta_{sd}^{LL}$ of order 
a few times $10^{-3}$ (see below), the Wilson coefficient 
$\Delta c_2$ can be significantly
larger than the value of the corresponding coefficient in the
Standard Model, which is proportional to $C_8$. 
This is interesting, since even in the Standard Model the 
contribution of $C_8$ to $\epsilon'/\epsilon$ is significant. 

\begin{table}
\centerline{\parbox{14cm}{\caption{\label{tab:1}
Values of the coefficients $\Delta c_i(m_c)$ and $\Delta\widetilde 
c_i(m_c)$ in units of $10^{-4}\,\mbox{Im}\,\delta_{sd}^{LL}$ and 
$10^{-4}\,\mbox{Im}\,\delta_{sd}^{RR}$, respectively, for common 
gluino and down-squark masses of 500\,GeV and different values of 
$m_{\tilde u_R}$. The last column shows the corresponding values in 
the Standard Model in units of $10^{-7}$.}}}
\vspace{0.3cm}
\begin{center}
\begin{tabular}{|l|rrr|r|}
\hline\hline
$m_{\tilde u_R}$\,[GeV] & 750 & 1000 & 1500 & SM \\
\hline
$\Delta c_1(m_c)$ & $-0.05$ & $-0.08$ & $-0.12$ & 0.42 \\ 
$\Delta c_2(m_c)$ & 2.12 & 3.19 & 4.16& $-1.90$ \\
$\Delta\widetilde c_3(m_c)$ & $-0.50$ & $-0.76$ & $-1.01$ & 20.64 \\ 
$\Delta\widetilde c_4(m_c)$ & 0.56 & 0.87 & 1.17 & $-7.63$ \\
\hline\hline
\end{tabular}
\end{center}
\vspace{-0.2cm}
\end{table}

In estimating the supersymmetric contribution to 
$\epsilon'/\epsilon$ we focus only on the $(V-A)\otimes (V+A)$ 
operators associated with the coefficients $\Delta c_1$ and 
$\Delta c_2$, because their matrix elements are chirally enhanced 
with respect to those of the $(V-A)\otimes(V-A)$ operators. 
The penguin operators 
contribute to the imaginary part of the isospin amplitude $A_2$. 
The real part is, to an excellent approximation, 
given by the matrix elements of the standard current--current 
operators in the effective weak Hamiltonian. Evaluating the matrix 
elements of the four-quark operators in the factorization 
approximation, and parametrizing nonfactorizable corrections by 
hadronic parameters $B_i^{(2)}$ as defined in \cite{Heff}, we 
obtain
\[
   \frac{\mbox{Im}\,A_2^{\rm susy}}{\mbox{Re}\,A_2}
   \approx \frac32\,\frac{m_K^2}{m_s^2(m_c)-m_d^2(m_c)}\,
   \frac{\mbox{Im}\,[\Delta c_2(m_c)+\frac13\Delta c_1(m_c)]\,
          B_8^{(2)}(m_c)}
        {|V_{us}^* V_{ud}|\,z_+(m_c)\,B_1^{(2)}(m_c)} \,.
\]
Following common practice we have neglected a tiny contribution 
proportional to the difference $B_7^{(2)}-B_8^{(2)}$. In the 
above formula $z_+$ is a combination of Wilson coefficients. The
product $z_+\,B_1^{(2)}=0.363$ is scheme independent and 
can be extracted from experiment. Note that at leading logarithmic 
order the scale dependence of the combination 
$\Delta c_2+\frac13\Delta c_1$ cancels the scale dependence 
of the running quark masses, 
and hence the hadronic parameter $B_8^{(2)}$ is scale independent.

\begin{figure}
\epsfxsize=10cm
\centerline{\epsffile{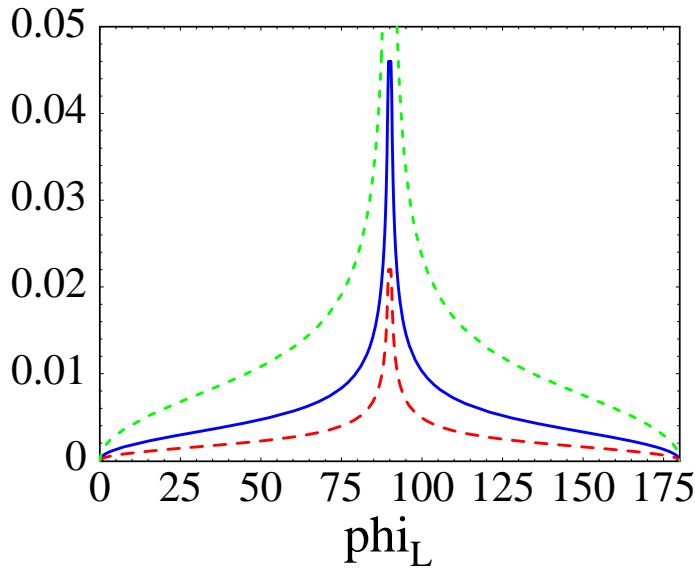}}
\vspace{0.3cm}
\centerline{\parbox{14cm}{\caption{\label{fig:bounds}
Upper bound on $|\mbox{Im}\,\delta_{sd}^{LL}|$ versus the weak phase 
$|\phi_L|$ (in degrees) for $m_{\tilde d_L}=500$\,GeV and 
$(m_{\tilde g}/m_{\tilde d_L})^2=1$ (solid), 0.3 (dashed) and 4
(short-dashed).}}}
\end{figure}

Putting everything together, we find for the supersymmetric
$\Delta I=\frac32$ contribution to $\epsilon'/\epsilon$
\begin{equation}\label{nice}
   \frac{\epsilon'}{\epsilon} \approx 19.2 
   \left[ \frac{500\,\mbox{GeV}}{m_{\tilde g}} \right]^2
   \left[ \frac{\alpha_s(\widetilde m)}{0.096} \right]^\frac{34}{21}
   \left[ \frac{130\,\mbox{MeV}}{m_s(m_c)} \right]^2\,
   B_8^{(2)}(m_c)\,X(x_d^L,x_u^R,x_d^R)\,
   \mbox{Im}\,\delta_{sd}^{LL} \,,
\end{equation}
where 
\[
   X(x,y,z) = \frac{32}{27}\,[f(x,y) - f(x,z)]
   + \frac{2}{27}\,[g(x,y) - g(x,z)] \,.
\]
The existence of this contribution requires a new CP-violating
phase $\phi_L$ defined by $\mbox{Im}\,\delta_{sd}^{LL}\equiv
|\delta_{sd}^{LL}|\sin\phi_L$. The measured values of $\Delta m_K$
and $\epsilon$ in $K$--$\bar K$ mixing give bounds on
$\mbox{Re}\,(\delta_{sd}^{LL})^2$ and 
$\mbox{Im}\,(\delta_{sd}^{LL})^2$, respectively, which 
can be combined to obtain a bound on
$\mbox{Im}\,\delta_{sd}^{LL}$ as a function of $\phi_L$. 
Using the most recent analysis of supersymmetric contributions to 
$K$--$\bar K$ mixing in \cite{long}, 
we show in Figure~\ref{fig:bounds} the results obtained for
$m_{\tilde d_L}=500$\,GeV and three choices of $m_{\tilde g}$. 

\begin{figure}
\epsfxsize=10cm
\centerline{\epsffile{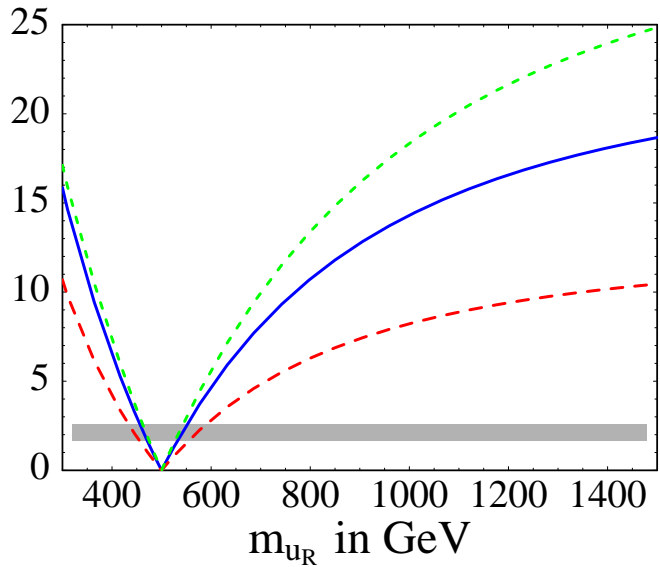}}
\vspace{0.3cm}
\centerline{\parbox{14cm}{\caption{\label{fig:eps}
Supersymmetric contribution to $|\epsilon'/\epsilon|$ (in units
of $10^{-3}$) versus $m_{\tilde u_R}$, for $m_s(m_c)=130$\,MeV,
$B_8^{(2)}(m_c)=1$, $m_{\tilde d_L}=
m_{\tilde d_R}=500$\,GeV, and $(m_{\tilde g}/m_{\tilde d_L})^2=1$ 
(solid), 0.3 (dashed) and 4 (short-dashed). The 
values of $|\mbox{Im}\,\delta_{sd}^{LL}|$ 
corresponding to the three
curves are 0.011, 0.005 and 0.027, respectively (see text). 
The band shows the average experimental value.}}}
\end{figure}

It is evident that the bound on $\mbox{Im}\,\delta_{sd}^{LL}$
depends strongly on the precise value of $\phi_L$. 
To address the issue of how large a supersymmetric contribution to 
$\epsilon'/\epsilon$ one can reasonably expect via the mechanism 
proposed in this Letter,
it appears unnatural to take the absolute maximum of the
bound given the peaked nature of the curves. To be
conservative we evaluate our result (\ref{nice}) taking for 
$\mbox{Im}\,\delta_{sd}^{LL}$
one quarter of the maximal allowed values shown in 
Figure~\ref{fig:bounds},
noting however that a larger effect could be obtained 
for the special case of $|\phi_L|$ very close to $90^\circ$.
Our results for $|\epsilon'/\epsilon|$ 
are shown in Figure~\ref{fig:eps} as a function of $m_{\tilde u_R}$ 
for the case
$m_{\tilde d_L}=m_{\tilde d_R}=500$\,GeV and the same three
values of $m_{\tilde g}$ considered in the previous figure. 
The choice $m_{\tilde d_L}=m_{\tilde d_R}$ is made for simplicity 
only and does not affect our conclusions in a qualitative way.
Except for the special case
of highly degenerate right-handed up- and down-squark masses, the 
$\Delta I=\frac32$ gluino box-diagram
contribution to $\epsilon'/\epsilon$ 
can by far exceed the experimental result, even taking into
account the bounds from $\Delta m_K$ and $\epsilon$. 
Indeed, even for moderate splitting Figure~\ref{fig:eps}
implies substantially stronger bounds on 
$|\mbox{Im}\,\delta_{sd}^{LL}|$ than those obtained from 
$K$--$\bar K$ mixing. This finding
is in contrast to the commonly held view that supersymmetric 
contributions to the 
electroweak penguin operators have a negligible impact 
on $\epsilon'/\epsilon$. 
In this context, it is worth noting 
that a large mass splitting between $\tilde u_R$ and $\tilde d_R$ 
can be obtained, e.g., in 
GUT theories without SU(2)$_R$ symmetry and 
with hypercharge embedded 
in the unified gauge group, without
encountering difficulties with naturalness \cite{Gian}.

The allowed contribution to 
$\epsilon'/\epsilon$ in Figure~\ref{fig:eps} increases with the 
gluino mass (for fixed squark masses) because  
the $K$--$\bar K$ bounds become weaker in this case.
If all supersymmetric masses are
rescaled by a common factor $\xi$, and the bounds 
from $K$--$\bar K$ mixing
are rescaled accordingly, the values for 
$\epsilon'/\epsilon$ scale like $1/\xi$ modulo logarithmic effects 
from the running coupling $\alpha_s(\xi\widetilde m)$. 
Therefore, even for larger squark masses of
order 1\,TeV the new contribution to $\epsilon'/\epsilon$ can
exceed the experimental value by a large amount, implying nontrivial
constraints on $\mbox{Im}\,\delta_{sd}^{LL}$.

Before concluding, we note that in the above discussion we have
made no assumption regarding the mass insertion parameter 
$\mbox{Im}\,\delta_{sd}^{RR}\equiv|\delta_{sd}^{RR}|\sin\phi_R$
for right-handed squarks. In models 
where $|\delta_{sd}^{RR}|$ is not highly suppressed 
relative to $|\delta_{sd}^{LL}|$, much tighter constraints
on $\mbox{Im}\,\delta_{sd}^{LL}$ can be derived by applying the
severe bounds on the 
product $\delta_{sd}^{LL}\,\delta_{sd}^{RR}$ obtained 
from the chirally-enhanced contributions to $K$--$\bar K$ mixing.
In analogy with Figure~\ref{fig:bounds}, we obtain an upper bound
on $|\mbox{Im}\,\delta_{sd}^{LL}|$ as a function of $\phi_L$ and 
$\phi_R$, which is sharply peaked along the line $\phi_L+\phi_R=0$
mod $\pi$ and scales like $|\delta_{sd}^{LL}/\delta_{sd}^{RR}|^{1/2}$. 
As above, we take one quarter of the peak value obtained 
using the results compiled in \cite{long}. Considering the
case $m_{\tilde d_L}=m_{\tilde g}=500$\,GeV for example, 
we find that the upper bounds on $|\mbox{Im}\,\delta_{sd}^{LL}|$ 
a reduced by a factor ranging from 3\% in the limit where 
$|\delta_{sd}^{RR}|=|\delta_{sd}^{LL}|$ to 8\% for 
$|\delta_{sd}^{RR}|=0.1|\delta_{sd}^{LL}|$. In the latter case,
the supersymmetric contribution to $\epsilon'/\epsilon$ can still be 
of order $10^{-3}$, i.e., comparable to the measured value. 
Moreover, significantly larger values can be obtained for special 
points in moduli space, where the weak phases obey 
$\phi_L+\phi_R\approx0$ mod $\pi$.

In summary, we have shown that in supersymmetric extensions of 
the Standard Model gluino box diagrams can yield a large 
$\Delta I=\frac32$ contribution to $\epsilon'/\epsilon$, which
only requires moderate mass splitting between the right-handed 
squarks, i.e., $(m_{\tilde u_R}-m_{\tilde d_R})/m_{\tilde d_R}>0.1$. 
In a large region of parameter space, the measured value of
$\epsilon'/\epsilon$ implies a significantly stronger bound on 
$\mbox{Im}\,\delta_{sd}^{LL}$ than is obtained from
$K$--$\bar K$ mixing.

\vspace{0.15cm}  
{\it Acknowledgments:\/} 
We are grateful to Yuval Grossman and JoAnne Hewett for useful 
discussions. A.K. is supported by the United States Department 
of Energy under Grant No.\ DE-FG02-84ER40153, and M.N. under 
contract DE--AC03--76SF00515.


\newpage
\begin{thebibliography}{99}

\bibitem{new}
A. Alavi-Harati et al.\ (KTeV Collaboration), \Prl{83}{1999}{22}.\\
A documentation of the NA48 result on $\epsilon'/\epsilon$ can be 
found at:\\
{\tt http://www.cern.ch/NA48}.

\bibitem{old}
G.D. Barr et al.\ (NA31 Collaboration), \plb{317}{1993}{233};\\
L.K. Gibbons et al.\ (E731 Collaboration), \Prl{70}{1993}{1203}.

\bibitem{updates}
M. Ciuchini, Nucl.\ Phys.\ B (Proc.\ Suppl.) {\bf 59}, 149 (1997);\\
S. Bertolini, M. Fabbrichesi and J.O. Eeg, \hepph{9802405};\\
S. Bosch et al., \hepph{9904408};\\
T.~Hambye, G.O. K\"ohler, E.A. Paschos and P.H. Soldan, 
\hepph{9906434};\\
A.A. Belkov, G. Bohm, A.V. Lanyov and A.A. Moshkin, \hepph{9907335}.

\bibitem{Gabb}
J. Hagelin, S. Kelley and T. Tanaka, \npb{415}{1994}{293};\\
E. Gabrielli, A. Masiero and L. Silvestrini, \plb{374}{1996}{80};\\
F. Gabbiani, E. Gabrielli, A. Masiero and L. Silvestrini, 
\npb{477}{1996}{321};\\
A.L Kagan, \Prd{51}{1995}{6196}.

\bibitem{SUSY}
A. Masiero and H. Murayama, \Prl{83}{1999}{907};\\
S. Khalil, T. Kobayashi and A. Masiero, \hepph{9903544};\\
S. Khalil and T. Kobayashi, \hepph{9906374};\\
K.S. Babu, B. Dutta and R.N. Mohapatra, \hepph{9905464};\\
S. Baek, J.-H. Jang, P. Ko and J.H. Park, \hepph{9907572};\\
R. Barbieri, R. Contino and A. Strumia, \hepph{9908255};\\
A.J. Buras et al., \hepph{9908371};\\
G. Eyal, A. Masiero, Y. Nir and L. Silvestrini, \hepph{9908382}.

\bibitem{Heff}
G. Buchalla, A.J. Buras and M.E. Lautenbacher, \Rmp{68}{1996}{1125}.

\bibitem{long}
M. Ciuchini et al., \jhep{9810}{1998}{008}.

\bibitem{Gian}
S. Dimopoulos and G.F. Giudice, \plb{357}{1995}{573}. 

\end{thebibliography}
\end{document}